\documentclass[aps,twocolumn,superscriptaddress]{revtex4}

\usepackage{graphicx,float}
\usepackage{epsfig}
\usepackage{epstopdf}

\usepackage{amsmath}
\usepackage{amssymb}
\usepackage{bm}
\usepackage{bbm}
\usepackage{dsfont}
\usepackage{bbold}
\usepackage{multirow}
\usepackage{color}
\usepackage[usenames,dvipsnames]{xcolor}
\usepackage[colorlinks=true,citecolor=blue,urlcolor=black]{hyperref}

\usepackage{hyperref}
\hypersetup{colorlinks=true,linkcolor=blue,citecolor=blue,urlcolor=blue}

\newcommand{\be}{\begin{equation}}
\newcommand{\ee}{\end{equation}}

\newcommand{\sket}[1]{{\ensuremath{\lvert#1\rangle}}}
\newcommand{\lket}[1]{{\ensuremath{\left\lvert#1\right\rangle}}}
\newcommand{\ket}[1]{\if@display\lket{#1}\else\sket{#1}\fi}

\newcommand{\sbra}[1]{{\ensuremath{\langle#1\rvert}}}
\newcommand{\lbra}[1]{{\ensuremath{\left\langle#1\right\rvert}}}
\newcommand{\bra}[1]{\if@display\lbra{#1}\else\sbra{#1}\fi}

\newcommand{\sbraket}[2]{{\ensuremath{\langle#1\rvert#2\rangle}}}
\newcommand{\lbraket}[2]{{\ensuremath{\left\langle#1\!\left\rvert\vphantom{#1}#2\right.\!\right\rangle}}}
\newcommand{\braket}[2]{\if@display\lbraket{#1}{#2}\else\sbraket{#1}{#2}\fi}

\newcommand{\sketbra}[2]{{\ensuremath{\lvert #1\rangle\!\langle #2\rvert}}}
\newcommand{\lketbra}[2]{{\ensuremath{\left\lvert #1\right\rangle\!\!\left\langle #2\right\rvert}}}
\newcommand{\ketbra}[2]{\if@display\lketbra{#1}{#2}\else\sketbra{#1}{#2}\fi}

\begin{document}

\title{Quantum key distribution with untrusted detectors}

\date{\today}

\author{P.~Gonz\'{a}lez}
\affiliation{Departamento de F\'isica, Universidad de Concepci\'on, 160-C Concepci\'on, Chile}
\affiliation{Center for Optics and Photonics, Universidad de Concepci\'on, Casilla 4016, Concepci\'on, Chile}
\affiliation{MSI-Nucleus for Advanced Optics, Universidad de Concepci\'on, Concepci\'on, Chile}

\author{L.~Reb\'on}
\affiliation{Instituto de F\'isica de La Plata, Universidad Nacional de La Plata, La Plata, Argentina}
\affiliation{Laboratorio de Procesado de Im\'{a}genes, Departamento de F\'{i}sica, Universidad de Buenos Aires, Buenos Aires, Argentina}

\author{T. Ferreira da Silva}
\affiliation{Optical Metrology Division, National Institute of Metrology, Quality and Technology,
 25250-020 Duque de Caxias, RJ, Brazil}

\author{M.~Figueroa}
\affiliation{Center for Optics and Photonics, Universidad de Concepci\'on,  Casilla 4016, Concepci\'on, Chile}
\affiliation{Departamento de Ingenier\'ia El\'ectrica, Universidad de Concepci\'on,160-C Concepci\'on, Chile}

\author{C.~Saavedra}
\affiliation{Departamento de F\'isica, Universidad de Concepci\'on, 160-C Concepci\'on, Chile}
\affiliation{Center for Optics and Photonics, Universidad de Concepci\'on,  Casilla 4016, Concepci\'on, Chile}

\author{M.~Curty}
\affiliation{EI Telecomunicaci\'on, Department of Signal Theory and Communications, University of Vigo, E-36310 Vigo, Spain}

\author{G.~Lima}
\affiliation{Departamento de F\'isica, Universidad de Concepci\'on, 160-C Concepci\'on, Chile}
\affiliation{Center for Optics and Photonics, Universidad de Concepci\'on,  Casilla 4016, Concepci\'on, Chile}
\affiliation{MSI-Nucleus for Advanced Optics, Universidad de Concepci\'on, Concepci\'on, Chile}

\author{G.~B.~Xavier}
\affiliation{Center for Optics and Photonics, Universidad de Concepci\'on,  Casilla 4016, Concepci\'on, Chile}
\affiliation{MSI-Nucleus for Advanced Optics, Universidad de Concepci\'on, Concepci\'on, Chile}
\affiliation{Departamento de Ingenier\'ia El\'ectrica, Universidad de Concepci\'on,160-C Concepci\'on, Chile}

\author{W.~A.~T.~Nogueira}

\affiliation{Center for Optics and Photonics, Universidad de Concepci\'on,  Casilla 4016, Concepci\'on, Chile}
\affiliation{MSI-Nucleus for Advanced Optics, Universidad de Concepci\'on, Concepci\'on, Chile}
\affiliation{Departamento de F\'isica, ICE, Universidade Federal de Juiz de Fora, Juiz de Fora, CEP 36036-330, Brazil}

\begin{abstract}
Side-channel attacks currently constitute the main challenge for quantum key distribution (QKD) to bridge theory with practice.
So far two main approaches have been introduced to address this problem,
(full) device-independent QKD and measurement-device-independent QKD. Here we present a third solution that might exceed
the performance and practicality of the previous two in circumventing detector side-channel attacks, which arguably
is the most hazardous part of QKD implementations. Our proposal has, however, 
one main requirement: the legitimate users of the system need to ensure that  
their labs do not leak any unwanted 
information to the outside. The security in the low-loss regime is guaranteed, while in the high-loss regime we already prove its robustness against some eavesdropping strategies. 
\end{abstract}

\maketitle

\section{Introduction}

Today quantum key distribution (QKD)~\cite{Gisin_RMP, Scarani_RMP, Lo_NP} faces the challenge of bridging the large gap between theory and practice. 
Theoretically, QKD offers perfectly secure communications 
based on the laws of physics. In practice, however, it does not because most physical devices do not operate as it is presumed 
in the security proofs. 
As a result, current QKD implementations suffer from security loopholes that allow for side-channel
attacks~\cite{sc01, sc02, sc03, sc04, sc05, blind1, blind2, scsource1, scsource2, scsource3, scsource4}.

To avoid these loopholes and recover the security of QKD realisations there are currently two main approaches where assumptions on the internal functioning of the measurement devices are avoided. The first one is called (full) device-independent QKD (diQKD)~\cite{diqkdrefsa1, diqkdrefsa2, diqkdrefsa3, diqkdrefsa4, time_reverse3}. Here, the legitimate users of the system (Alice and Bob) treat
their apparatuses as two ``black boxes''. Given that certain conditions are satisfied, it is 
possible to prove the security of diQKD based solely on the violation of a Bell inequality. Importantly, this solution 
can remove all side-channels from the quantum part of a QKD implementation.  
Its main drawback, however, is that it requires a loophole-free Bell test~\cite{bell1, bell2, bell3, bell4, bell5, bell6} with distant communicating parties, which is yet to be achieved. Also, its expected secret key rate with current technology
is very low at practical distances~\cite{Gisin2010a, Gisin2010b}.

The second approach is called measurement-device-independent QKD (mdiQKD)~\cite{mdiQKD}.
In contrast to diQKD, Alice and Bob need to know their state preparation processes but they can treat the 
measurement device as a ``black box'' fully controlled by the eavesdropper (Eve). 
This solution eliminates all side-channels from the measurement unit, which can be regarded as the weakest part 
of a QKD implementation~\cite{sc01, sc02, sc03, sc04, sc05, blind1, blind2}, and guarantees a very high performance.   
Indeed, mdiQKD tolerates a high optical loss of more than $40$ dB and it can give 
a secret key rate similar to that of standard entanglement-based QKD protocols~\cite{ent_qkd}. 
Moreover, its feasibility has already been proven both in laboratories
and via field-tests~\cite{Rubenok1, Rubenok2, Rubenok3, Rubenok4, Rubenok5, Rubenok6}. This suggests the viability of mdiQKD to connect theory and practice in QKD.
This approach has, however, two slight drawbacks. First, mdiQKD requires high-visibility two-photon 
interference using two different light sources, which makes its experimental implementation more demanding than 
that of conventional QKD systems. Second, 
the current finite-key security bounds~\cite{finite_mdi} require relatively large post-processing data block sizes 
to achieve good 
performance. 

Here we propose an alternative solution to remove detector side-channels in QKD realisations.
It follows a similar spirit to that of mdiQKD. That is, Alice and Bob need to characterise their state
preparation processes but do not have to trust the measurement device, which is treated as a ``black box''.
Note, however, that the concept of ``black box'' is now different from that of mdiQKD. 
In particular, we requiere that Alice and Bob know the optical elements contained in the box, but no knowledge is required on the way they work or on which quantum system they operate \cite{footnote}.
This is so to prevent attacks that exploit the fact that Eve can build the measurement unit herself and she includes additional elements that leak key information to the channel~\cite{barrett, Bing}.
Indeed, our proposal requires that Alice and Bob guarantee that the measurement system does not leak any unwanted information to the outside (just like in diQKD). 
This could be achieved, in principle, by placing the measurement apparatus within Bob's laboratory, and with a measurement device built by them, albeit not necessarily characterised \cite{Bing}. 
This condition can be checked/fulfilled in most practical scenarios. In this case, the only relevant information to prove security is the statistics of the input and output data from the box. 

In doing so, as will be explained below, it is possible to avoid the problem of interfering photons from independent sources, which 
considerably simplifies its experimental implementation when compared to mdiQKD. In the low-loss regime, the security of our approach is guaranteed by the results in~\cite{cyril}. Here we also conjectured its security in the high-loss regime by analysing a particular class of attacks. In parallel to this work, further developments towards more general security proofs of single-photon two-qubit device-independent QKD  \cite{Bing, Ma} have been carried out, together with other experimental implementations  \cite{CLim, Pan, Li}.

\section{Description of the protocol}

\begin{figure}[t]
\centerline{\includegraphics[width=0.49\textwidth]{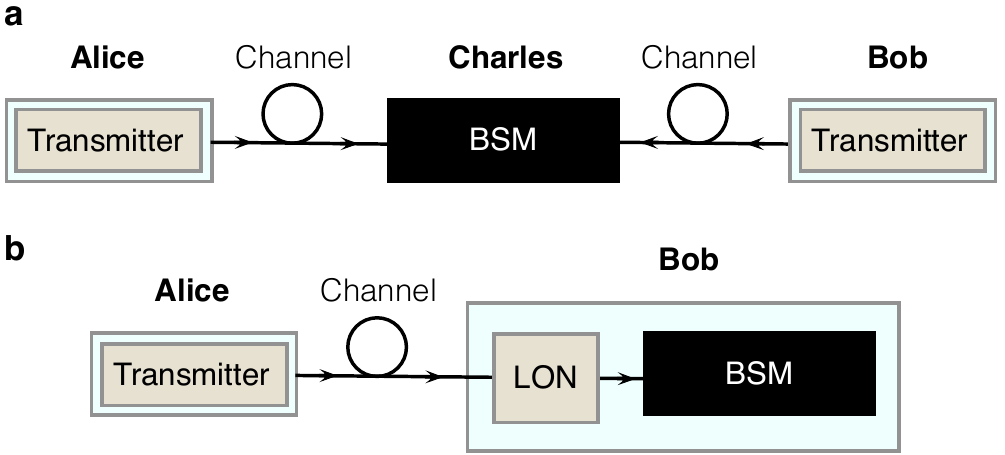}}
\vspace{-0.2cm}
\caption{(a) Schematic diagram of measurement-device-independent QKD (mdiQKD)~\cite{mdiQKD}. Alice and Bob 
prepare different quantum states and send them to an untrusted relay Charles, which can be treated 
as a black box fully controlled by Eve. Charles is supposed to implement a Bell state measurement (BSM) that projects the 
incoming signals into a Bell state. 
(b) Schematic diagram of our proposal. Alice generates different quantum states and sends them
to Bob. On receiving the signals, Bob encodes his information by means of a 
trusted linear optics network (LON), which can
be regarded as Bob's transmitter (when compared to mdiQKD). This LON does not include any light source but it simply
manipulates the state of the incoming signals.  
Afterwards, Bob implements a BSM, which is treated as a black box. 
 In the figure: (brown box) characterised device; (black box) uncharacterised device; and
(light turquoise box) secure lab, {\it i.e.}, the lab does not leak any unwanted information to the outside.  
}\label{fig_scheme}
\end{figure}

The key idea of our protocol is illustrated in Fig.~\ref{fig_scheme}. For comparison, this figure also includes
a schematic diagram of mdiQKD~\cite{mdiQKD}.  
Alice uses a transmitter to prepare different quantum states that she sends 
to Bob. On the receiving side, Bob uses a linear optics network (LON) to manipulate 
the state of the incoming signals. Alice's transmitter and Bob's LON are both trusted and characterised.
When compared to mdiQKD, this LON can be regarded as 
Bob's trusted transmitter, although it does not include any light source. 
Afterwards, Bob is supposed to implement a Bell state measurement (BSM), which
is considered to be a black box. 

One requirement for the measurement device, as mentioned above, is that no unwanted information leaks from the BSM. In practice, a simple way to achieve this is to place the measurement device within Bob's shielded lab, and that Bob builds the BSM himself, such that he can assure that it does not contain any eavesdropping device prepared by Eve \cite{Bing}. This is indeed the expected situation in most realistic scenarios. Even though Bob builds the BSM, he does not 
need to characterise the exact functioning of the optical elements within the BSM ({\it e.g.}, polarisation 
rotators, beamsplitters, single-photon detectors, etc). That is, in the security analysis one can treat the whole 
BSM as a black box, where the only relevant information is the input and output data of the box. 

Let us describe our quantum key distribution protocol by using a particular 
example of a possible implementation. This setup is schematically shown in
Fig.~\ref{fig1}. The protocol can be summarised with the following three steps:

\begin{figure}[t]
\centerline{\includegraphics[width=0.488\textwidth]{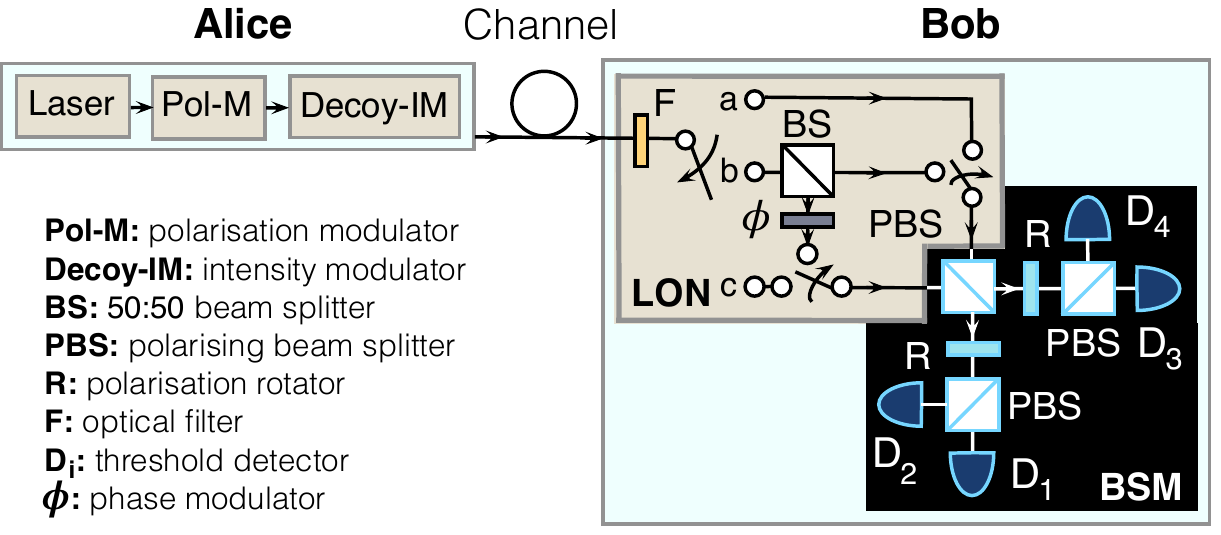}}
\vspace{-0.2cm}
\caption{Schematic diagram of an example of a possible implementation of our method. 
Alice generates
phase-randomised 
weak coherent pulses (WCPs) in different BB84 polarisation states~\cite{bb84}, and sends them to Bob. In addition, she prepares 
decoy-states~\cite{decoy1, decoy2, decoy3} using
an intensity modulator (Decoy-IM). Bob employs a trusted LON to encode his information on the incoming signals by using their path 
degree of freedom. For this, he uses 
an optical switch that sends the arriving states through one out of three possible optical paths of the same length (paths $a$, $b$ and $c$ in the figure). 
Two additional optical switches are used to guarantee that the selected path is actually connected to the 
polarising beamsplitter (PBS).
The switches are represented in the figure with the standard drawing of arrows and white connectors.
The phase modulator $\phi$ shifts the phase of each pulse by $0$ or $\pi$. 
Bob measures the outcoming pulses from the LON with a linear-optics single-photon BSM~\cite{Kim_PRA_2003}. 
A successful BSM result is obtained when only one detector $D_i$ ``clicks''.
The polarisation rotator $R$ changes the horizontal (vertical) polarisation to a $45^\circ$ ($-45^\circ$) linear polarisation.
As in Fig.~\ref{fig_scheme}: (brown box) characterised device; (black box) uncharacterised device; and
(light turquoise box) secure lab.  
}\label{fig1}
\end{figure}

{\it Step 1:}
Alice sends Bob phase-randomised weak coherent pulses (WCPs), together with decoy signals~\cite{decoy1, decoy2, decoy3}, prepared in different BB84 polarisation states~\cite{bb84}. 
For each signal, these states are selected independently and at random
from two mutually-unbiased bases, {\it e.g.}, either a rectilinear ($H$ [horizontal] or $V$ [vertical]) or a diagonal ($45^\circ$ or $-45^\circ$) polarisation basis. 

{\it Step 2:}
On receiving the transmission, Bob employs a LON to encode his information on the incoming pulses using their path 
degree of freedom. For this, he utilises 
an optical switch that distributes the arriving signals into one out of three possible optical paths
($a$, $b$ and $c$ in Fig.~\ref{fig1}), which he selects independently and at random for each pulse.
In analogy to Alice, we shall denote the paths $a$ and $c$ in Fig.~\ref{fig1} as Bob's rectilinear bases, 
and the path $b$ (with $\phi=0$ or $\phi=\pi$) as Bob's diagonal bases. 
Moreover, we will consider that Bob's bit value associated to selecting the path $a$ ($c$) is equal to that of Alice when she chooses $H$ ($V$) polarisation,
and that Bob's bit value associated to selecting the path $b$ with $\phi=0$ ($\phi=\pi$) is equal to that of Alice when she employs 
$45^\circ$ ($-45^\circ$) polarisation. If we compare this procedure to that of mdiQKD, 
one could say that to select path $a$ ($c$) in our proposal is somehow equivalent to Bob preparing a 
$H$ ($V$) polarisation
signal in mdiQKD, and similarly for the diagonal basis. Once Bob has encoded his information, 
 the signals are recombined at a polarising beamsplitter (PBS) and then measured with a 
 linear-optics single-photon BSM~\cite{Boschi98, Kim_PRA_2003}. 
A successful BSM result corresponds to observing a ``click'' in only one detector $D_i$, with $i\in\{1, \ldots, 4\}$. If two or more detectors click simultaneously, the event is considered unsuccessful. 

{\it Step 3:} 
Alice and Bob employ an authenticated
classical channel to announce their results. In particular, Bob declares which pulses produced a successful BSM result together with the 
Bell state obtained. Also, Alice and Bob broadcast the polarisation and path basis that they have used
to generate and measure each successful signal respectively. 
They keep the data associated
with those successful events where they used the same basis and discard the rest. 
In addition, they use the decoy-state method~\cite{decoy1, decoy2, decoy3} to estimate the yield ({\it i.e.}, the probability that Bob obtains a successful BSM result) and
the quantum bit error rate (QBER) for various $n$-photon states.
Like in mdiQKD, the key point is that with this information Alice and Bob can determine whether or not the BSM is working well enough to be able to distill a secret key.
If this is the case,
either Alice or Bob applies a bit flip to part of her/his data to 
assure
that their bit strings are correctly correlated (see Table~\ref{table1}). 
They then finally perform error-correction and privacy amplification procedures to obtain a final secret key.
\begin{table}[htb]
  \centering
  \begin{tabular}{lcccc}
    \quad & \multicolumn{4}{c}{``Clicking" detector}\\
    \hline\hline
  Alice \& Bob & \quad $D_1$ \quad & \quad $D_2$ \quad & \quad $D_3$ \quad & \ \quad $D_4$ \ \quad \\
  \hline
  {\rm Rectilinear basis} & \ \quad\quad - \ \quad\quad & \quad - \quad  & \quad {\rm Bit flip} \quad  & \quad {\rm Bit flip}  \quad  \\
  {\rm Diagonal basis} & \ \quad\quad - \ \quad\quad  & \quad {\rm Bit flip} \quad & \quad - \quad & \quad {\rm Bit flip} \quad \\
 \hline\hline
\end{tabular}
\caption{To guarantee that their bit strings are correctly correlated,
either Alice or Bob applies a bit flip to part of her/his data, depending on which detector $D_i$ ``clicked" (which identifies the 
Bell state obtained by the BSM)
and the basis setting selected. Detections at D$_1$, D$_2$, D$_3$ and D$_4$ correspond to the Bell state projections $|\phi^+\rangle$, $|\phi^-\rangle$, $|\psi^+\rangle$ and $|\psi^-\rangle$, where the Bell states are written in terms of the hybrid polarization-path encoding \cite{Boschi98, Kim_PRA_2003}.}\label{table1}
\end{table}

Let us emphasise that the method described above could be applied as well to other QKD protocols 
like, for instance, the three-state scheme~\cite{three2, three3}. Also, it could be adapted to other encoding strategies
({\it e.g.},
phase encoding or time-bin encoding). In addition, 
let us point out that the use of optical switches (within Bob's 
LON) is not essential; indeed, it is possible to 
design alternative receivers without these elements.

\section{Security assumptions}

Before we analyse the security of the protocol, let us begin by stating the security assumptions. In particular, we assume that Alice and Bob have access to
(i) true random number generators, (ii) trusted classical post-processing techniques and (iii), an 
authenticated classical channel, (iv) Alice's source and Bob's LON 
are fully characterised and cannot be influenced by Eve, and (v) 
Alice's and Bob's labs do not leak any unwanted information 
to the outside. 

The first three assumptions are also required in conventional QKD systems. The fourth one needs special attention.
In principle, it is reasonable to expect that Alice can verify the states she sends to Bob
in a fully protected environment outside Eve's control. For this, she could 
protect herself with different optical elements like, 
for instance, optical isolators, 
optical filters and a monitoring detector; also, she could use random sampling techniques. This is precisely the  
scenario we face in mdiQKD. 
The case of Bob, however,
is more delicate. 
This is so because he actually receives signals from the quantum channel.
Eve may try to perform, for example, a so-called Trojan horse attack \cite{Trojan,Trojan2}.
That is, she could launch bright light pulses into Bob's LON and then analyse the back-reflected light. 
In doing so, Eve could try to determine Bob's bit value ({\it i.e.}, 
the position of his optical switch in the example above) for each arriving signal. 
In practice, however, 
this type of attacks (or similar ones) might be avoided as well
by including additional components on Bob's side, just like in the case of Alice. 
For example, 
Bob could insert several optical circulators to attenuate the back-reflected light
together with optical filters to remove undesired modes
and a monitoring detector to test the incoming and/or reflected
light. Further details on possible countermeasures against Trojan horse attacks can be found in~\cite{Trojan,Trojan2}.

Alternatively, Eve could also try to manipulate the correct operation of Bob's LON 
by shifting, for instance, the frequency or the arrival time of the incoming pulses. 
This way she might influence the functioning of both 
the beamsplitter and the phase modulator within the LON. 
Again, however, in practice one expects that Bob could avoid such type of attacks by using, for example, optical filters together with 
a time-dependent attenuator. This attenuator could restrict the arrival time of the signals 
to only a certain time window where the devices work as predicted by the mathematical models used to prove security. In the example given by Fig.~\ref{fig1} 
the role of such attenuator is performed
by the optical switch. Furthermore, note that
Bob could even remove the 
phase modulator within his LON. If Alice sends him only three 
different states, it can be shown that 
this scenario ({\it i.e.}, without phase modulator on Bob's side) would be completely equivalent to that of the 
three-state protocol~\cite{three2, three3}. According to the results in~\cite{three3}
the expected performance in this case
would be basically the same as that of the original situation where Alice and Bob use four 
different states. 

To conclude this part, let us discuss the 
fifth assumption considered. Note that this assumption is also required both 
in diQKD~\cite{barrett} and mdiQKD. The only difference is that in mdiQKD this condition does not affect the 
measurement unit, which can be located outside Alice's and Bob's secure labs. 
In our proposal, however,  Bob's state preparation process is 
performed by his LON, which is situated between the channel and the BSM. Therefore, 
if we treat the BSM as a ``black box'' under Eve's control
and, moreover, this box can send any information that Eve wishes to the 
outside, Eve might try to learn the whole key without introducing any errors, as discussed in \cite{Bing}. For this reason, it is essential that Bob can guarantee 
the requirement
that the BSM does not leak unwanted information 
to the outside.

\section{Security analysis}

We now evaluate the security of the protocol. From the results in~\cite{cyril} we have 
that our scheme is secure against general attacks
in the low-loss regime ({\it i.e.}, 
when the overall transmittance of the single-photon pulses sent by Alice 
is greater or equal to $65.9\%$) given that Bob's measurement device is memoryless.
This is so because the work in~\cite{cyril} contains our proposal as a special case; more precisely, 
its security analysis considers the worst-case scenario where Bob's device is untrusted. 
For this reason, such result, although it guarantees security when the loss is low, 
might be over pessimistic 
since here we assume that part of Bob's apparatus ({\it i.e.}, his LON) can be actually trusted.  

Below we conjecture the security of our scheme also in the practical and relevant scenario of high losses.  
For this, we prove its security against a certain class of attacks. In particular, 
we assume that Eve can block or correlate the single-photon pulses sent by Alice with an ancilla system
in her hands, 
but she cannot add additional photons to these pulses. That is, whenever Alice 
emits a single-photon signal Bob receives either vacuum or a single-photon. In addition, we 
permit 
that Eve can decide the output of the BSM for each pulse sent by Alice. A full security proof 
against general attacks in the high-loss regime is left for future analysis. Note that recently, a similar mathematical proof to the one presented here, with an equivalent level of security, was given in \cite{CLim}.

We use similar arguments to those employed in mdiQKD~\cite{mdiQKD}, which relies
on the security of a time reversed EPR-based QKD protocol~\cite{time_reverse1, time_reverse2, time_reverse3}. Indeed, it can
be shown that the protocol illustrated in Fig.~\ref{fig1}, when viewed in the reverse order, is equivalent to a
counter-factual entanglement based BB84 protocol~\cite{ent_based}. That is, 
whenever Bob observes  a single ``click" in a detector $D_i$ in our protocol, this corresponds to the situation
where Eve distributes a certain Bell state $\ket{\phi_i}$ in the counter-factual protocol.

To see this, we focus on the
single-photon states sent by Alice.
In an equivalent virtual protocol, her signal state preparation can be thought of as follows.
Alice prepares an entangled bipartite state of the form
$\ket{\Psi}_{AA'}=\sum_i\sqrt{p_i}\ket{a_i}_{A}\ket{\psi_i}_{A'}$. If she measures the virtual 
system $A$ in the orthonormal basis $\ket{a_i}_{A}$, she effectively prepares the
BB84 states $\ket{\psi_i}_{A'}$ with probability $p_i$. Moreover, she can also
incorporate in her virtual measurement the information about the reduced density matrix of 
system $A$, {\it i.e.}, $\rho_{A}={\rm Tr_{A'}(\ket{\Psi}_{AA'}\bra{\Psi})}$,
which is known and fixed by the state preparation process \cite{curty1, curty2}. 

The case of Bob is more subtle.
In a virtual protocol, he first prepares the virtual state $\ket{\Phi}_{B}=\sum_i\sqrt{p_i}\ket{b_i}_{B}$, with
$\ket{b_i}_{B}$ being an orthonormal basis for system $B$. Then, whenever he receives
a single-photon signal $\sigma_{A'}$ from the channel, which might have been manipulated by Eve, he applies a controlled unitary operation
$U_{BA'}=\sum_i \ket{b_i}\bra{b_i}_{B}\otimes{}U_{i,A'}$ on systems $B$ and $A'$, where the unitary operators
$U_{i,A'}$ are fixed by his state preparation process ({\it i.e.}, by the action of his LON). That is, each operator
$U_{i,A'}$ corresponds to one particular setting of his 
optical switch and phase modulator. Now, the key point is that it
can be shown (see the Appendix) 
that after applying $U_{BA'}$ the reduced density matrix of system $B$, that we denote as $\rho_{B}$, 
is fixed
and equal to $\rho_{A}$ {\it independently} of the incoming single-photon state $\sigma_{A'}$. That is, 
Bob's virtual system $B$ is in the same state as if he would have followed the same state preparation process as Alice
to generate BB84 signals. Now, the scenario is precisely the same as that of mdiQKD. That is, in the virtual picture  
Alice and Bob could in principle keep their systems $A$ and $B$ in a quantum
memory and delay their measurements on them until the 
BSM is completed. In such virtual scenario the protocol is then directly equivalent to an entanglement based
BB84 scheme~\cite{ent_qkd, ent_based}.

As a result, we have that the asymptotic secret key rate formula has the form $R\geq \sum_i \max\{R_i,0\}$, with $R_i$ denoting the key rate associated with those events 
where Bob observes a ``click" only in detector $D_i$. This
parameter is given by~\cite{key_formula1, key_formula2, key_formula3}
\begin{equation}\label{seckey}
R_i\geq q\big\{p_0Y_{i,0}+p_1Y_{i,1}[1-h(e_{i,1})]-Q_if(E_i)h(E_i)\big\}.
\end{equation} Here, the coefficient $q$ denotes the efficiency of the protocol ({\it i.e.}, $q=1/2$ for the
standard BB84 protocol~\cite{bb84} and $q\approx 1$ for its
efficient version~\cite{efbb84}); $p_n=\exp{(-\mu)}\mu^n/n!$ is the probability that Alice
sends Bob a signal which contains $n$ photons,
with $\mu$ being the average photon number of the signals;
$Y_{i,n}$ denotes the conditional probability that Bob only observes a ``click" in detector $D_i$
given that Alice sent him an $n$-photon state; the parameter $e_{i,n}$ represents the QBER
of those $n$-photon signals which only produce a click
in detector $D_i$; $h(x)=-x\log_2(x)-(1-x)\log_2(1-x)$ denotes the binary Shannon entropy function;
the term $Q_i$ represents the probability that Bob only obtains a ``click" in detector $D_i$
when Alice sends him a signal state, {\it i.e.}, $Q_i=\sum_n p_nY_{i,n}$; the parameter
$E_i$ is the overall QBER associated with a detection in $D_i$, {\it i.e.}, $E_i=\sum_n p_nY_{i,n}e_{i,n}/Q_i$;
and $f(x)$ is an inefficiency function for the error correction process in the protocol
(typically $f(E_i)\geq1$; with the Shannon limit $f(E_i)=1$).

Equation~(\ref{seckey}) contains three parameters
which are not directly observed in the experiment:
$Y_{i,0}$, $Y_{i,1}$ and $e_{i,1}$. To estimate these quantities
we use the decoy-state method~\cite{decoy1, decoy2, decoy3}. Here, for simplicity,
we consider that Alice employs an infinite number of decoy settings and, therefore,
Alice and Bob are able to obtain the precise values of these parameters.
In the practical scenario where Alice and Bob only use a finite number
of decoy settings one can solve such estimation problem either by using 
linear programming tools, or by employing, for instance, the analytical procedure reported in
Ref.~\cite{finite_decoy}. 

As a final remark, let us emphasise, once again, that in the scenario where
Eve can replace the single-photons with multi-photon or strong pulses, the
security of our scheme is not yet clear. However, one possible solution
might be to ensure that the overall detection efficiency of all outputs of
the measurement device are the same \cite{Bing, thiago}. This could be verified
through the input/output statistics, and finely tuned by Bob through
optical attenuators in each output. Another alternative solution might be to
follow the ideas introduced in \cite{CLim} and assume that the linear optical
elements within the BSM are also trusted (i.e., to consider that the only
untrusted components within Bob are the detectors). Both approaches,
nevertheless, require further investigations.

\section{Simulation}

For simulation purposes we consider inefficient and noisy threshold detectors $D_i$,
and we use experimental parameters from Ref.~\cite{param}. In addition, for simplicity,
we assume that all detectors are identical and their
dark counts are, to a good approximation, independent of the incoming signals.
Moreover, we use a channel
model that includes an intrinsic error rate of $1.5\%$,
simulating the misalignment and instability of the optical system.

\begin{figure}[t]
\centerline{\includegraphics[width=0.45\textwidth]{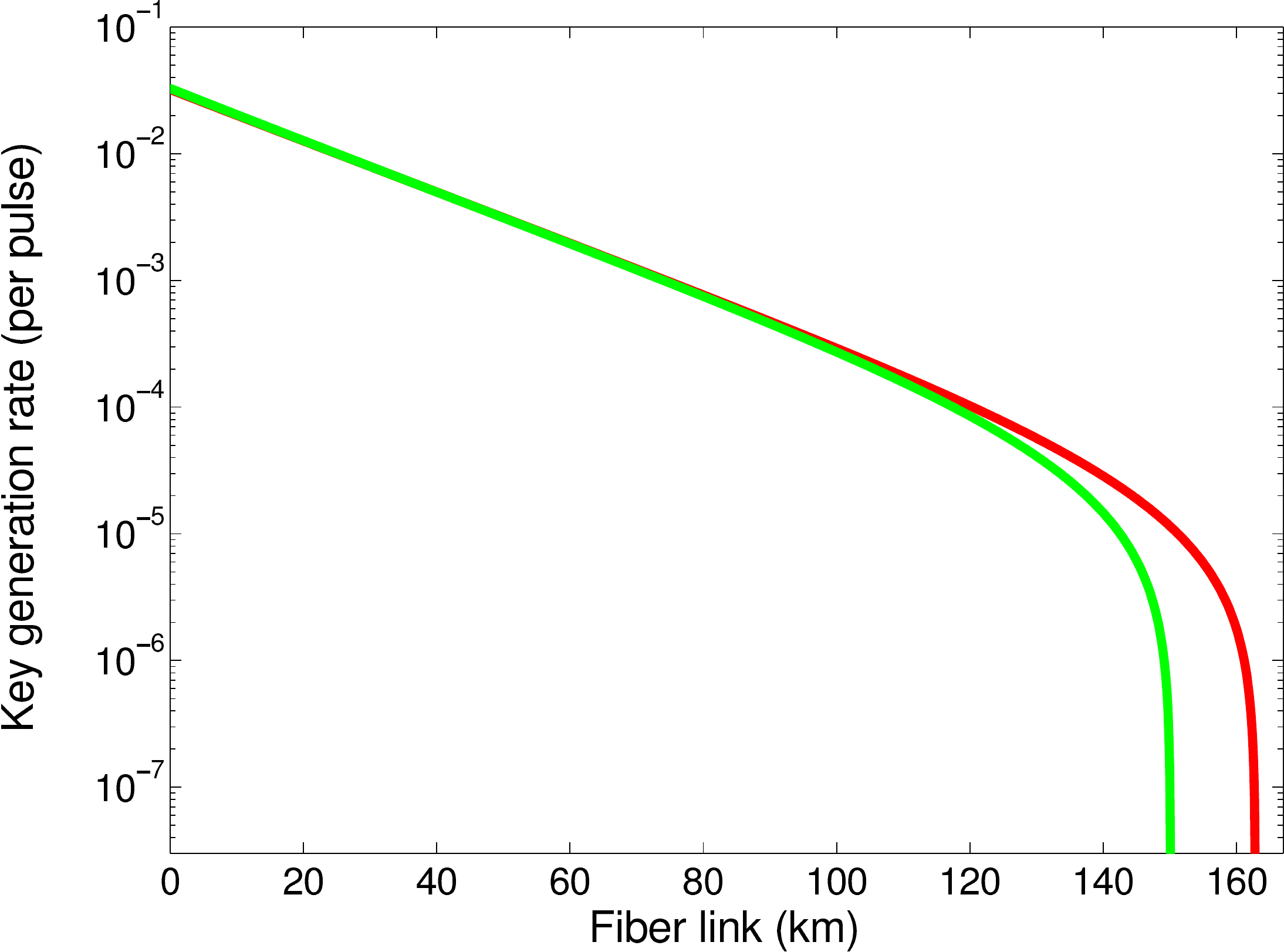}}
\vspace{-0.2cm}
\caption{Lower bound on the secret key rate $R$ given by Eq.~(\ref{seckey}) in logarithmic scale for the setup illustrated in
Fig.~\ref{fig1} with WCPs (green curve). For simulation purposes, we use experimental parameters from Ref.~\cite{param}: 
the loss coefficient of the quantum channel is $0.2$ dB/km,
the intrinsic error rate due to misalignment and instability of the optical system is $1.5\%$,
the overall detection efficiency of the detectors $D_i$ is $14.5\%$, and the background count rate is $6.02\times{}10^{-6}$.
Furthermore, we consider that the parameter $q\approx 1$ \cite{efbb84} and the efficiency of the error correction protocol
satisfies $f(E_i)=1.16$. For comparison, this figure also includes a lower bound on the secret key rate for a standard
decoy-state BB84 system with an infinite number of decoy settings and an active measurement setup
(red curve)~\cite{decoy1, decoy2, decoy3}. }\label{fig_rate}
\end{figure}

The resulting lower bound on the secret key rate $R$ given by Eq.~(\ref{seckey}) is illustrated in Fig.~\ref{fig_rate}.
For a given total system loss, {\it i.e.}, including the losses in the channel and in Bob's detection apparatus,
we optimise the lower bound on $R$ over the average photon number $\mu$ of Alice's signal states,
which is around $0.7$ for most of the distances.
For comparison, this figure also includes a lower bound on the secret key rate for an asymptotic
decoy-state BB84 system with an infinite number of decoy settings and an active receiver with two detectors~\cite{decoy1, decoy2, decoy3}. We consider the BB84 scheme with two-detectors as a comparison because this is a standard configuration for this protocol, whereas in our new proposal four detectors are required
to maximise its key rate. As a result, we find that both secret key rates are very similar. Only the cutoff point of the standard decoy-state BB84 scheme ($163$ km) is slightly larger than that of the protocol illustrated
in Fig.~\ref{fig1} ($150$ km). This is because in the case of the standard decoy-state BB84 system Bob's measurement device
has a lower overall dark count rate than that of our proposal, since it only contains two detectors instead of four. 
Most importantly, our scheme illustrated in Fig.~\ref{fig1} delivers a secret key rate which is approximately
two orders of magnitude higher than that of mdiQKD (please see Fig. 2 in Ref.\cite{mdiQKD}) for the experimental parameters considered, although now the covered distance is shorter.

\section{Proof-of-principle experiment}

For the sake of completeness, we performed a proof-of-principle experiment to simply demonstrate that all the required states by the protocol can be successfully generated and detected.
More specifically, a complete key exchange session with realistic security requirements (that is, random generation and measurement by Alice and Bob, decoy states, error correction and privacy amplification) is outside the scope of the current work.  
Also, for simplicity, instead of using phase-randomised WCPs,
the signal states emitted by Alice are generated with a continuous wave laser at $690$ nm attenuated to the single photon level, calibrated to a detection window of 4 ns.
Alice controls the polarisation of these signals with a half-wave plate (HWP),
and sends them to Bob through a free-space channel. Bob's measurement device is a
slightly modified version of that illustrated in Fig.~\ref{fig1}. In particular, the rectilinear path basis
is defined by blocking one of the two possible paths of the interferometer, while
the diagonal path basis follows the description of Fig.~\ref{fig1}. This simpler configuration is equivalent to the one depicted in
 Fig.~\ref{fig1}, but now the rectilinear basis suffers an additional $3$dB loss. Note, however,
 that one could remove this extra loss (introduced by the rectilinear path basis) if Alice and Bob resort to the diagonal and circular bases. In this case, Bob would always have the interferometer with both arms unblocked, and use four different phase settings, two for each basis. 

As a phase modulator, we use a mirror mounted over a piezoelectric actuator in one of the paths of the interferometer.
No active stabilisation of the interferometer was needed for the time-scale involved in the experimental measurements, which were taken with an integration time of 1s per data point. In order to implement the BSM we employ two HWPs set to $22.5^{\circ}$ as rotators $R$. The detectors $D_i$ are commercial pigtailed single-photon detectors based on Si avalanche photodiodes, operating in free-running mode. The overall raw visibility of the interference curves was $88.4 \pm 0.2\%$.

\begin{figure*}[t]
\centerline{\includegraphics[width=0.9\textwidth]{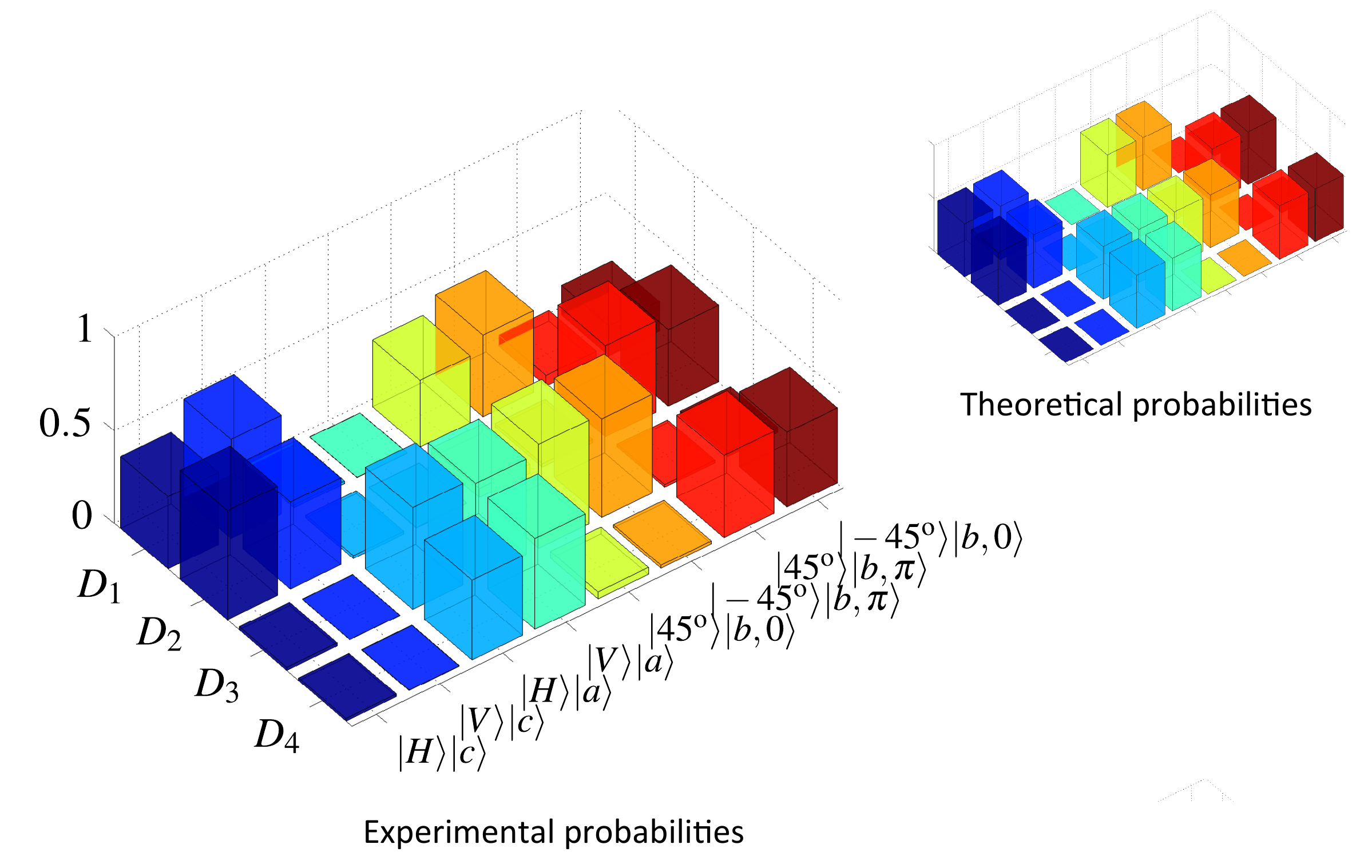}}
\caption{Experimental and theoretical (inset) probabilities
to obtain a click in detector $D_i$ for all possible combinations of states when both Alice and
Bob use compatible bases. See Table \ref{table2} for further details. The states 
$\ket{a}$ and $\ket{c}$ in the figure denote the paths a and c respectively, while $\ket{b,0}$ and $\ket{b,\pi}$ represent path b with 
$\phi=0$ and $\phi=\pi$ respectively.}\label{fig3}
\end{figure*}

We experimentally measured all possible combinations of states used for the BB84 protocol when both Alice and
Bob simultaneously choose the rectlinear or diagonal bases. The single counts are recorded simultaneously on all four detectors using independent counting circuits programmed on FPGA-based electronics. The results are shown in Fig.~\ref{fig3} (see also Table \ref{table2}). They are in good agreement with the theoretical predictions. From the measured visibility, the average projected QBER over all different states is $5.8 \pm 0.1\%$.

\begin{table*}[htb]
  \centering
  \begin{tabular}{lcccc}
    \quad & \multicolumn{4}{c}{Detection probabilities}\\
    \hline\hline
 State & \quad $D_1$ \quad & \quad $D_2$ \quad & \quad $D_3$ \quad & \ \quad $D_4$ \ \quad \\
  \hline
 \ket{H}\ket{c} & \ \quad\quad $0.3863 \pm 0.0007$  \ \quad\quad & \quad $0.5823 \pm 0.0010$ \quad  & \quad $0.0168 \pm 0.0001$ \quad  & \quad $0.0146 \pm 0.0001$  \quad  \\
 \ket{V}\ket{c} & \ \quad\quad $0.5316 \pm 0.0008$  \ \quad\quad & \quad $0.4652 \pm 0.0007$ \quad  & \quad $0.0010 \pm 0.00002$ \quad  & \quad $0.0022 \pm 0.00003$  \quad  \\
 \ket{H}\ket{a} & \ \quad\quad $0.0152 \pm 0.0001$  \ \quad\quad & \quad $0.0153 \pm 0.0001$ \quad  & \quad $0.5432 \pm 0.0009$ \quad  & \quad $0.4263 \pm 0.0007$  \quad  \\
 \ket{V}\ket{a} & \ \quad\quad $0.0010 \pm 0.00002$  \ \quad\quad & \quad $0.0020 \pm 0.00004$ \quad  & \quad $0.5101 \pm 0.0008$ \quad  & \quad $0.4862 \pm 0.0008$  \quad  \\
 \ket{45^{\circ}}\ket{b,0} & \ \quad\quad $0.3574 \pm 0.0005$  \ \quad\quad & \quad $0.0488 \pm 0.0001$ \quad  & \quad $0.5569 \pm 0.0007$ \quad  & \quad $0.0369 \pm 0.0001$  \quad  \\
 \ket{-45^{\circ}}\ket{b,\pi} & \ \quad\quad $0.4357 \pm 0.0005$  \ \quad\quad & \quad $0.0175 \pm 0.00008$ \quad  & \quad $0.5309 \pm 0.0007$ \quad  & \quad $0.0158 \pm 0.00007$  \quad  \\
 \ket{45^{\circ}}\ket{b,\pi} & \ \quad\quad $0.0569 \pm 0.0001$  \ \quad\quad & \quad $0.4871 \pm 0.0006$ \quad  & \quad $0.0144 \pm 0.00006$ \quad  & \quad $0.4417 \pm 0.00001$  \quad  \\
  \ket{-45^{\circ}}\ket{b,0} & \ \quad\quad $0.1591 \pm 0.0002$  \ \quad\quad & \quad $0.4091 \pm 0.0005$ \quad  & \quad $0.0285 \pm 0.000009$ \quad  & \quad $0.4033 \pm 0.00005$  \quad  \\

 \hline\hline
\end{tabular}
\caption{Detection probabilities for the eight possibilities when Alice and Bob use compatible bases.}\label{table2}
\end{table*}

\section{Conclusions}

We have presented a new approach to the problem of detector side-channels in
practical QKD, which arguably constitutes the Achilles heel of current experimental realisations. 
It builds on the approach known as measurement-device-independent QKD (mdiQKD)~\cite{mdiQKD}. However,
when compared to mdiQKD, it has two main potential advantages. First, it is simpler to implement experimentally 
since it does not require interference of independent laser sources, just like conventional QKD systems. This means, in particular, that 
no active tracking of the arrival times of independent photons nor frequency control of their sources are necessary.
Also, it does not need coincidence detections, which is particularly important for setups with low overall detection efficiency.
Second, although in this paper we have assumed for simplicity
the asymptotic scenario where Alice sends Bob an infinite number of signals, one expects that 
the finite secret key rate of our approach will be much higher than that of mdiQKD as now only Alice needs to send decoy states. 
For the same reason, one also expects that the size of the post-processing data blocks will be significantly smaller than those required in 
mdiQKD, which is essential in practice \cite{finite_decoy_1}. 

In this work, we already prove the security of our scheme against general attacks in the low-loss regime and against a particular class of attacks in the high-loss regime. Nevertheless, in order for it to be a plausible alternative to mdiQKD, it is crucial to demonstrate 
its security against general attacks also in the high-loss regime. This important open question is left for further studies.   

\section*{Acknowledgements}

We would like to thank Koji Azuma, 
Hoi-Kwong Lo, Kiyoshi Tamaki, Bing Qi and Zhen-Qiang Yin for very useful discussions. L. Reb\'on thanks the Center for Optics and Photonics (Universidad de Concepci\'on) for hospitality and CONICET for financial support. This work was supported by the grants FONDEF Idea IT13I10017 CONICYT PFB08-024, Milenio P10-030-F, FONDECYT (grants No. 11110115, 1150101, 1120067 and 1151278),
and by the Galician Regional Government (program ``Ayudas para proyectos de investigaci\'on desarrollados por investigadores emergentes'', and 
consolidation of Research Units: AtlantTIC), and the Spanish Government (project TEC2014-54898-R).  In addition W. A. T. Nogueira and P. Gonz\'{a}lez thank CNPq (Brazil) and Conicyt (Chile) respectively for financial support. 

\section*{Appendix}

\noindent{\bf Reduced density matrix $\rho_B$.}
Here we briefly show that after applying the controlled unitary operation $U_{BA'}$
the reduced density matrix $\rho_B$ of Bob's virtual system is equal to that of Alice's virtual system. 

For this, we first obtain an expression 
for the unitary operators $U_{i,A'}$, with $i\in\{1, \ldots, 4\}$. As explained in the main text, here we will assume that system $A'$ is a qubit. 
In particular, when Bob selects path $a$ we have that
\begin{eqnarray}\label{eq_late}
U_{1,A'}\ket{1,0}_{A'}\ket{0,0}_{\rm aux}&=&\ket{1,0}_{\rm inp_1}\ket{0,0}_{\rm inp_2}, \nonumber \\ 
U_{1,A'}\ket{0,1}_{A'}\ket{0,0}_{\rm aux}&=&\ket{0,1}_{\rm inp_1}\ket{0,0}_{\rm inp_2},
\end{eqnarray} 
where the state $\ket{1,0}$ denotes one photon in horizontal polarisation
and $\ket{0,1}$ is one photon in vertical polarisation.
System $\rm aux$ represent the signal in the orthogonal path ({\it i.e.}, in path $c$). The labels inp$_1$ and inp$_2$
denote, respectively, the signals in the two input ports of the BSM. That is, Eq.~(\ref{eq_late}) tells us that when the single-photon $A'$ is prepared in horizontal (vertical)
polarisation, and Bob selects path $a$, then we have one photon in horizontal (vertical) polarisation in the input port inp$_1$ of the BSM. 

When Bob chooses path $c$ we have that
\begin{eqnarray}\label{eq_l}
U_{2,A'}\ket{1,0}_{A'}\ket{0,0}_{\rm aux}&=&\ket{0,0}_{\rm inp_1}\ket{1,0}_{\rm inp_2}, \nonumber \\ 
U_{2,A'}\ket{0,1}_{A'}\ket{0,0}_{\rm aux}&=&\ket{0,0}_{\rm inp_1}\ket{0,1}_{\rm inp_2},
\end{eqnarray}
where system $\rm aux$ denotes again the signal in the orthogonal path ({\it i.e.}, in path $a$ in this case). That is, 
when the single-photon $A'$ is prepared in horizontal (vertical)
polarisation, and Bob selects path $c$, then we have one photon in horizontal (vertical) polarisation in the input port inp$_2$. 

Using the same procedure we obtain that $U_{3,A'}$, which corresponds to selecting 
path $b$ and $\phi=0$, and  $U_{4,A'}$ (for path $b$ and $\phi=\pi$) have the form
\begin{eqnarray}\label{eq_l}
U_{3,A'}\ket{1,0}_{A'}\ket{0,0}_{\rm aux}&=&1/\sqrt{2}\big[\ket{1,0}_{\rm inp_1}\ket{0,0}_{\rm inp_2}\nonumber \\ 
&&+\ket{0,0}_{\rm inp_1}\ket{1,0}_{\rm inp_2}\big], \nonumber \\ 
U_{3,A'}\ket{0,1}_{A'}\ket{0,0}_{\rm aux}&=&1/\sqrt{2}\big[\ket{0,1}_{\rm inp_1}\ket{0,0}_{\rm inp_2}\nonumber \\
&&+\ket{0,0}_{\rm inp_1}\ket{0,1}_{\rm inp_2}\big], \nonumber \\
U_{4,A'}\ket{1,0}_{A'}\ket{0,0}_{\rm aux}&=&1/\sqrt{2}\big[\ket{1,0}_{\rm inp_1}\ket{0,0}_{\rm inp_2}\nonumber \\ 
&&-\ket{0,0}_{\rm inp_1}\ket{1,0}_{\rm inp_2}\big], \nonumber \\ 
U_{4,A'}\ket{0,1}_{A'}\ket{0,0}_{\rm aux}&=&1/\sqrt{2}\big[\ket{0,1}_{\rm inp_1}\ket{0,0}_{\rm inp_2}\nonumber \\ 
&&-\ket{0,0}_{\rm inp_1}\ket{0,1}_{\rm inp_2}\big].
\end{eqnarray}

System $\sigma_{A'}$ can always be written as $\sigma_{A'}=\sum_i q_i \ket{\phi_i}_{A'}\bra{\phi_i}$ for certain pure states
$\ket{\phi_i}_{A'}$. This means, in particular, 
that in order to prove that $\rho_B=\rho_A$ for any input state $\sigma_{A'}$ it is enough to show that this 
condition is satisfied for any signal $\ket{\phi_i}_{A'}=\alpha\ket{1,0}_{A'}+\beta\ket{0,1}_{A'}$. 
Let 
\begin{eqnarray}
\ket{\phi}_{B,\rm inp_1,\rm inp_2}&=&U_{BA'}\sum_i\sqrt{p_i}\ket{b_i}_{B}\ket{\phi_i}_{A'}\ket{0,0}_{\rm aux} \nonumber \\
&=&\sum_i\sqrt{p_i}\ket{b_i}_{B}U_{i,A'}\ket{\phi_i}_{A'}\ket{0,0}_{\rm aux}.
\end{eqnarray}
Then, we have that $\rho_B={\rm Tr}_{\rm inp_1,\rm inp_2}(\ket{\phi}_{B,\rm inp_1,\rm inp_2}\bra{\phi})$. 
Finally, by combining the equations above now it is straightforward to show that, independently of the state 
$\ket{\phi_i}_{A'}$, 
 indeed $\rho_B$ is a density matrix of rank two equal to 
$\rho_A={\rm Tr_{A'}(\ket{\Psi}_{AA'}\bra{\Psi})}$ with 
$\ket{\Psi}_{AA'}=\sum_i\sqrt{p_i}\ket{a_i}_{A}\ket{\psi_i}_{A'}$ and where 
$\ket{a_i}_{A}$
in an orthonormal basis and $\ket{\psi_i}_{A'}$ denotes de BB84 single-photon states. 
We omit this step here for simplicity, but the calculations are direct.
Using the 
same type of calculations it can also be shown that $\rho_B$ is basis independent.



\begin{thebibliography}{000}
\bibitem{footnote}
Note, however, that this condition is not needed in mdiQKD where the measurement device can be even manufactured by Eve.
\end{thebibliography}
\end{document}